\documentclass[useAMS,usenatbib]{mn2e}

\usepackage[dvipdfmx]{graphicx}
\usepackage{amsmath,amssymb,natbib,siunitx,color,booktabs}
\usepackage{color}

\input epsf
\usepackage{graphicx}
\newcommand{\beq}{\begin{equation}}
\newcommand{\beqa}{\begin{eqnarray}}
 \newcommand{\eeq}{\end{equation}}
\newcommand{\eeqa}{\end{eqnarray}}

\newcommand{\lmk}{\left(}
\newcommand{\rmk}{\right)}
\newcommand{\lnk}{\left\{ }
\newcommand{\rnk}{\right\} }

\newcommand{\so}{M_\odot}
\newcommand{\mch}{{\cal M}}

\begin{document}
%
%
%

\title[]{Potential Tertiary Effects on the LISA Verification Binary HM Cancri}

\author[]
{Naoki Seto\\
Department of Physics, Kyoto University, Kyoto 606-8502, Japan
}

\date{\today}
\maketitle

%
%
%
%
%
%
\begin{abstract}
 Two groups recently analyzed the long-term orbital  evolution of HM Cancri, which is one of the most important verification binaries for the space gravitational wave detector LISA.  By using the reported first and second derivatives of its orbital frequency $f$, we discuss potential tertiary effects on this binary. We found that, in contrast to the first derivative $\dot f$,  the second derivative $\ddot f$ might be strongly affected by a dark tertiary component such as an old white dwarf with an outer orbital period of $\sim$250\,years.

\end{abstract}


 \section{introduction}
 
HM Cancri (RX J0806.3+1527) is  a mass-transferring white dwarf binary of the orbital period of 321sec (corresponding to the orbital frequency of $f=3.1$mHz), identified about 20 years ago \citep{2002MNRAS.332L...7R,2002A&A...386L..13I}. 
A circular binary like HM Cancri emits a nearly monochromatic  gravitational wave (GW) at the frequency corresponding to twice the orbital frequency.  Among the known Galactic compact binaries, HM Cancri has the highest GW frequency of $2f=6.2$mHz \citep{2018MNRAS.480..302K,2022arXiv220306016A}.  It is also considered to be one of the brightest verification binaries for space GW detectors such as LISA \citep{2022arXiv220306016A}, Taiji \citep{2018arXiv180709495R} and TianQin \citep{2016CQGra..33c5010L}.  Indeed, the Chinese project TianQin will have an orbital configuration for optimally detecting  HM Cancri \citep{2016CQGra..33c5010L,2020PhRvD.102f3021H}.  These space GW detectors are expected to { individually resolve} $\sim 10^4$ close white dwarf  binaries in our Galaxy \citep{2022arXiv220306016A}.

Shortly after the identification of HM Cancri, the variation rate  $\dot f$  of its orbital frequency $f$ was measured at ${\dot f}\sim3.6 \times 10^{-16}\,{\rm Hz \,s^{-1}}$ (\citealt{2005ApJ...627..920S} see also \citealt{2004MSAIS...5..148I}).  This result was statistically unexpected, since a mass transferring white dwarf binary is {considered to}  stay mostly in outspiral state (${\dot f}<0$, see e.g., \citealt{2004MNRAS.350..113M}).
At present,  some other interacting white dwarf binaries  (e.g., V407Vul and SDSSJ0651) are known to be  in inspiral state ${\dot f}>0$ \citep{2018MNRAS.480..302K}.

Quite recently, for HM Cancri,   respectively  using its long-term  X-ray and optical data, two groups   measured the second frequency derivative $\ddot f$  and reported  ${\ddot f}\sim -10^{-26}{\rm Hz\,s^{-2}}$ \citep{2021ApJ...912L...8S,2023MNRAS.518.5123M}.    Interestingly, this numerical value is different from a preceding theoretical expectation ${\ddot f}\sim 10^{-28}{\rm Hz\,s^{-2}}$ (\citealt{2006ApJ...649L..99D}) with respect to both the sign and the magnitude. 
 
 As pointed out by \cite{2023MNRAS.518.5123M},  the observed second derivative $\ddot f$  might indicate that HM Cancri is at a rare evolutionary stage shortly before the frequency maximum and was discovered as a result of a selection effect (see also \citealt{2021ApJ...912L...8S}).  In this paper, as a potential mechanism for affecting the observed rate $\ddot f$, we discuss the gravitational perturbation by a tertiary component around the binary.  In fact, for  Galactic binaries detectable with LISA, there are many theoretical studies on the GW phase modulation  induced by their tertiaries (see e.g., \citealt{2018PhRvD..98f4012R,2021MNRAS.502.4199X} and references therein). In view of these activities, the recent $\ddot f$ measurements will provide us with a unique opportunity to actually examine LISA sources in the context of tertiary perturbation.

 In this paper, we study the potential tertiary effects in the following order.  In \S 2, we summarize the observed long-term orbital evolution of HM Cancri. 
 In \S 3, we examine the apparent orbital phase modulation induced  by a  tertiary and make a case study for HM Cancri. We discuss related aspects in \S  4. \S 5 is devoted to a  short summary.

 \section{observed orbital evolution}
 
 \cite{2021ApJ...912L...8S} analyzed the X-ray data of HM Cancri from Chandra and NICER with a baseline of $\sim20$ years.  He 
  fitted  its long-term orbital phase evolution  using a cubic function  
\begin{equation}
\Phi(t)= 2\pi \lmk f t+\frac{{\dot f}t^2}{2!}+\frac{{\ddot f}t^3}{3!}   \rmk+\phi  \label{wf}
\end{equation}
with the time origin $t=0$  at a certain epoch in January  2004 and the initial orbital phase constant $\phi$.  The fitted  first and second  frequency derivatives are 
\beq
{\dot f}=(3.557\pm  0.005)\times 10^{-16} {\rm Hz\, s^{-1}},\label{ch1}
\eeq
\beq
{\ddot f}=(-8.95\pm1.4)\times 10^{-27}{\rm Hz\,s^{-2}}. \label{dds}
\eeq
Here the error bars show the $1\sigma$ uncertainties. 

\cite{2023MNRAS.518.5123M} made an  orbital timing analysis  for HM Cancri, using its optical data accumulated (more evenly) in the past $\sim 20$ years.  They found a good fit with the cubic functional form (\ref{wf}).  The resultant second derivative ${\ddot f}=(-5.38\pm2.1)\times 10^{-27}{\rm Hz\,s^{-2}}$ is somewhat different from the X-ray result (\ref{dds}) (but nearly overlapping error bars).  At present, among LISA's verification binaries, we know the second derivative $\ddot f$  only for HM Cancri.

The intrinsic  frequency evolution of an interacting white dwarf  binary is mainly determined by the competition between the gravitational radiation reaction and the mass transfer \citep{1967AcA....17..287P,2004MNRAS.350..113M}. Generally speaking, the first derivative $\dot f$  can be  measured more easily,   with a shorter time baseline.  In fact, as for HM Cancri, results similar to the value (\ref{ch1})  were reported soon after its discovery \citep{2004MSAIS...5..148I,2005ApJ...627..920S}. 

If the observed rate $\dot f$ is dominated by the radiation reaction, we  have ${\dot f}\sim  {\dot f}_{\rm GW}\propto  f^{11/3}\mch^{5/3}$ with the chirp mass $\mch$.   On the basis of this relation, the chirp mass of HM Cancri can be estimated to be $\mch\sim {0.32}\so$ \citep{2021ApJ...912L...8S}.  Similarly, if the system is controlled by the radiation reaction,  the second derivative $\ddot f$ will  be close to 
\beq
{\ddot f}_{\rm GW}\simeq\frac{11{\dot f}^2}{3f}\sim 1.5\times 10^{-28} {\rm Hz\,s^{-2}},\label{gwf}
\eeq
as  suggested by  \cite{2006ApJ...649L..99D} well ahead of  the actual measurement of  $\ddot f$ by \cite{2021ApJ...912L...8S} and \cite{2023MNRAS.518.5123M}.  In reality, the observed rate  (\ref{dds}) is totally different from the expected rate  ${\ddot f}_{\rm GW}$ above.

\cite{2023MNRAS.518.5123M} examined the fitting results $\{{\dot f},{\ddot f}  \}$ with simulation models based  on the Modules for Experiments in Stellar Astrophysics (MESA)  code \citep{2019ApJS..243...10P}.  As mentioned earlier, they argued that  HM Cancri might be at an evolutionary state  shortly ($\sim1000$yr) before the frequency maximum $\dot f=0$ (see also \citealt{2007ApJ...655.1010G}  for oscillation of the frequency  $f$ due to spin-orbit coupling). They also pointed out  that HM Cancri might be discovered as a result of a selection effect  related to the time dependence of the mass transfer rate.

 \section{tertiary perturbation}

\subsection{Tertiaries around Galactic LISA  sources}

As in the case of short-period main-sequence star binaries \citep{2006A&A...450..681T}, observations indicate that a significant fraction of white dwarf binaries have tertiary components \citep{2017A&A...602A..16T}.  On   the theoretical side,  dynamical  processes such as the Kozai-Lidov mechanism (see e.g., \citealt{2016ARA&A..54..441N,2023arXiv230613130S}) might be relevant for the earlier evolutionary stages of some ultra-compact binaries. 

As discussed in the literature  (see e.g., \citealt{2018PhRvD..98f4012R,2021MNRAS.502.4199X} and references therein),  tertiary component could modulate the GW phase of a white dwarf binary which is  detectable with space GW detectors.   In an optimistic  case, LISA might detect a Jupiter mass planet around a white dwarf binary \citep{2008ApJ...677L..55S,2019NatAs...3..858T}.  In a pessimistic case, a tertiary might just become  a noise source at decoding intrinsic binary evolution from GW data  \citep{2018PhRvD..98f4012R,2021MNRAS.502.4199X}. 

By examining existing information on verification binaries, we can make a prior research for the future space GW detectors. Given the recent measurements of the second derivative $\ddot f$  for HM Cancri, it would be timely to conduct a tertiary study with its real data  ({see also \citealt{2023arXiv230704453S} for the measurability of  the observed rate $\ddot f$ with LISA}).

As an outer tertiary component around HM Cancri, to be consistent with current observation,  we conservatively suppose an underluminous object that will be outshined by the inner accreter  and  difficult to be detected with electromagnetic telescopes.  While the distance to HM Cancri  from the Earth has  large uncertainties  \citep{2023MNRAS.518.5123M}, an old white dwarf will  be a suitable candidate for such a tertiary.

 \subsection{Orbital Phase Modulation}
 
 Here we briefly discuss the outer tertiary perturbation on the apparent inner orbital evolution. We can find related references on the pulsar timing analysis  (see e.g., \citealt{2016ApJ...826...86K,2016MNRAS.460.2207B}). 
  We put the apparent orbital phase $\Phi(t)$ by 
 \begin{equation}
\Phi(t)=\Phi_{\rm int}(t)+\Phi_{\rm mod}(t). \label{ph1}
\end{equation}
 The  term $\Phi_{\rm int}(t)$ represents the intrinsic inner binary   evolution  and is assumed to be well described by
 \begin{equation}
\Phi_{\rm int}(t)= 2\pi \lmk f_{\rm int} t+\frac{{\dot f}_{\rm int}t^2}{2!}+\frac{{\ddot f}_{\rm int}t^3}{3!}   \rmk+\phi  \label{wf2}
\end{equation}
with the expansion coefficients $\lnk f_{\rm int},{\dot f}_{\rm int},{\ddot f}_{\rm int}  \rnk$ defined at $t=0$. 
In Eq. (\ref{ph1}), the  term  $\Phi_{\rm mod}(t)$ originates from  the modulation of  the inner binary barycenter due to the tertiary.  In terms of the radial distance $D(t)$ to the inner barycenter, we have
 \begin{equation}
\Phi_{\rm mod}(t)=-2\pi f_{\rm int} D(t) c^{-1}.
\end{equation}
In fact, considering the light travel time  between the  inner binary and the  observer,  the apparent orbital phase should be better modeled by $\Phi(t)=\Phi_{\rm int}[t-D(t)/c]$. However, if the  outer orbital period  $P$ (the characteristic  timescale for the variation of $\dot D$) is shorter than the inner evolution time $|f_{\rm int}/{\dot f}_{\rm int}|$ and $|f_{\rm int}/{\ddot f}_{\rm int}|^{1/2}$,  our expression (\ref{ph1})  (without the couplings like ${\dot  f}D$)  is a good approximation.

 Assuming  that the outer orbital period $P$ is much longer than the observation time $T\sim20\,$yr, the projected distance $D$ can be efficiently Taylor expanded  with the derivative coefficients $\{{\dot D},{\ddot D},{\dddot D}\}$ defined at $t=0$.  We then have 
 \begin{equation}
{ f}=\left. \frac1{2\pi} \frac{d\Phi}{dt}\right|_{t=0}= { f}_{\rm int}- f_{\rm int} {\dot D}c^{-1}
\
\end{equation}
 \begin{equation}
{\dot f}=\left. \frac1{2\pi} \frac{d^2\Phi}{dt^2}\right|_{t=0}= {\dot f}_{\rm int}- f_{\rm int} {\ddot D}c^{-1}  \label{f1}
\end{equation}
 \begin{equation}
{\ddot f}=\left. \frac1{2\pi}\frac{d^3\Phi}{dt^3}\right|_{t=0}= {\ddot f}_{\rm int}-f_{\rm int} {\dddot D}c^{-1} \label{f2}
\end{equation}
for the apparent inner orbital phase $\Phi(t)$.

Here
we define the notations $\Delta {\dot f}$ and  $\Delta {\ddot f}$ for the correction terms in Eqs. (\ref{f1}) and (\ref{f2}) and call them  the acceleration and jerk terms   respectively.
Given $|{\dot D}/c|\ll 1$, we can practically use $f_{\rm int}=f$ for these term as
\begin{equation}
\Delta {\dot f}=-f  {\ddot D}c^{-1},~~\Delta {\ddot f}=-f  {\dddot D}c^{-1}. \label{aj}
\end{equation}
Note that  the constant bulk velocity of the triple system is unimportant for our study.

 \subsection{Circular Orbit}
 Now we  assume that the outer orbit is circular with the semimajor axis $R$ and the inclination angle $I$.  We put  the tertiary mass by $m_3$ and the total mass of the triple system by $M_T$. Then we can put the projected distance $D(t)$ by
 \begin{equation}
D(t)=A\cos(2\pi t/P+\varphi)
\end{equation}
with the orbital phase $\varphi$ at $t=0$.  The amplitude $A$ is given by 
 \begin{equation}
A= \frac{   m_3\sin I}{M_T}R=FR
\end{equation}
with the factor
$F\equiv  (m_{3}/M_T)\sin I <1$.   From the Kepler's law,  we have the outer  orbital period as     
 \beqa
P&=& 2\pi \lmk  \frac{R^3}{GM_T}\rmk^{1/2}\\
& =&250 \lmk\frac{ M_T}{2\so}\rmk^{-1/2} \lmk \frac{R}{{\rm 50\, au}}\rmk^{3/2}{\rm yr}.
\eeqa
 Here we ignored the corrections associated with the time variation of the radial projection vector (see e.g., \citealt{1970SvA....13..562S,1992RSPTA.341...39P}). For a target  system at a Galactic distance $\gtrsim 1$kpc, they are much  smaller than the reported values in Eqs. (\ref{ch1}) and (\ref{dds}) \citep{2023MNRAS.518.5123M}. From Eq.  (\ref{aj}), we have the acceleration and jerk terms as 
 \begin{equation}
\Delta {\dot f}=  f A (2\pi/P)^2 \cos\varphi \label{ac2}
\end{equation}
 \begin{equation}
\Delta{\ddot f}= - f A (2\pi/P)^3 \sin\varphi. \label{ac22}
\end{equation}
Their magnitudes are given by 
 \beqa
f A (2\pi/P)^2&=&5.0  \times 10^{-17}F\lmk \frac{f}{\rm 3.1mHz}  \rmk \lmk \frac{M_T}{2\so}  \rmk^{1/3} \nonumber   \\
 & &\times\lmk \frac{P}{\rm 250 yr}  \rmk ^{-4/3}{\rm Hz\, s^{-1}} \label{fa1} \\
f A (2\pi/P)^3&=&4.0  \times 10^{-26}F\lmk \frac{f}{\rm 3.1mHz}  \rmk  \lmk \frac{M_T}{2\so}  \rmk^{1/3} \nonumber \\
& &\times \lmk \frac{P}{\rm 250 yr}  \rmk ^{-7/3} {\rm Hz\, s^{-2}}.\label{fa2}
\eeqa

\begin{table}
\caption{ The observed values  and the potential tertiary effects for $\dot f$ and $\ddot f$. The tertiary effects are given  for $(P,F)=(250\,{\rm yr},0.5)$ and $(100\,{\rm yr},0.05)$ with the fixed total mass of $M_T=2.0M_\odot$ (see Eqs. (\ref{fa1}) and  (\ref{fa2})).  We omit the phase factors $\cos \varphi$ and $-\sin\varphi$ presented in  Eqs.  (\ref{ac2}) and (\ref{ac22}). }
 \centering
 
\begin{tabular}{@{}l|lll@{}}
\toprule
    &    observed values  & (250yr,0.5) & (100yr,0.05)  \\ \midrule
 ${\dot f}~~({\rm Hz\, s^{-1}})$ &  $3.6\times 10^{-16}$  &  $2.5\times 10^{-17}$ & $8.5\times 10^{-18}$       \\
 
${\ddot f}~~({\rm Hz\, s^{-2}})$ &$-9.0\times 10^{-27}$ & $ 2.0\times 10^{-26}$  &   $ 1.7\times 10^{-26}$       \\

 \bottomrule
\end{tabular}
\end{table}

\subsection{Case Study for HM Cancri}
We now make  case studies for HM Cancri. Here we need to pay attention to both amplitude and the timescale of the tertiary perturbation. 

\subsubsection{Outspiral to Inspiral}
Our first question is whether the acceleration term $\Delta {\dot f}$ can change  an outspiral state ${\dot f}_{\rm int}<0$ to the observed inspiral state ${\dot f}>0$ (see also \citealt{2021MNRAS.502.4199X}). If this is the case, we  have 
\beq
\Delta {\dot f}={\dot f}-{\dot f}_{\rm int}>{\dot f}.
\eeq
From Eqs. (\ref{dds})  and (\ref{fa1}), the outer orbital period $P$ should be comparable to the observational time $T\sim20$yr.  
Then the apparent inner frequency  $f(t)$ should show a strong modulation pattern, covering a considerable fraction of the outer orbital  period $P$. This contradicts to the actual data (in particular the optical ones) that are well fitted by the cubic model with a small parameter $\ddot f$ (relative to Eq. (\ref{fa2}) for $P\ll 250$yr). 

Note that the difficulty will not be resolved, even with a stellar mass black hole of  $m_3\sim 10\so$. For an outer orbital period $P\gg T\sim20$yr, we will simply have   $|\Delta {\dot f}|\ll {\dot f}\simeq {\dot f}_{\rm int}$, and the acceleration term $\Delta {\dot f}$ will be unimportant for HM Cancri.

We have assumed that outer  orbit is circular. However, even for an  eccentric orbit, it will be generally  difficult to suitably generate the observed value $\dot f>0$ from an outspiral state ${\dot f}_{\rm int}<0$. This is because both   the orbital variation timescale and the magnitude of the acceleration are mainly determined by the distance between the inner binary and the tertiary. 

\subsubsection{Correction to $\ddot f$}
 The situation is largely different for the observed second derivative  $\ddot f$ presented in Eq. (\ref{dds}). For example, taking $F=0.5$ and $P\sim250$\,yr$(\gg T\sim20$\,yr), we can make $|\Delta {\ddot f}|\gtrsim |{\ddot f}|$ (see Table 1).
Thus, even from a small intrinsic value ${\ddot f}_{\rm int}\simeq {\ddot f}_{\rm GW}$ (see Eq. (\ref{gwf})), the jerk term $\Delta  {\ddot f}$ can generate the observed rate $\ddot f= {\ddot f}_{\rm int}+\Delta  {\ddot f}$ with a tertiary mass $m_3\sim1\so$. In future, with a sophisticated technology,   we might find an electromagnetic counterpart of the tertiary, depending on its properties.  

 For $P\sim100$\,yr, an underluminous M-dwarf  or just  a brown dwarf can be an effectual  perturber with $F\sim 0.05$.   For such a relatively short  orbital period $P$, we might observe a systematic  deviation from the cubic   time fitting (\ref{wf}) before LISA's launch.

\section{discussion}
So far, we have concentrated on HM Cancri, which  has the highest orbital frequency $f$ among the known  verification binaries for LISA.  In addition, at present,  it is the unique verification binary with a measured second derivative $\ddot f$.   Therefore, it was natural to initiate our  study from HM Cancri.  Below, we discuss potential  tertiary analysis for other LISA sources.

In many cases, compared with the second  derivative ${\ddot f}$,  the first derivative $\dot f$ can be measured with a shorter time  baseline.  In fact, as mentioned earlier, the first derivatives $\dot f$ have been measured for some of verification binaries (see e.g., \citealt{2023arXiv230313573B} for a recent result).  Note that the acceleration term $\Delta {\dot f}$ in Eq. (\ref{ac2})  is proportional to the  inner orbital frequency $f_{\rm int}(\simeq f)$. In contrast,  the magnitude of the intrinsic rate $|{\dot f}_{\rm int}|$ typically decreases more rapidly at the lower  frequency  regime,  reflecting the nature of gravitational radiation reaction.  Therefore, in the relation ${\dot f}={\dot f}_{\rm int}+\Delta {\dot f}$,  the acceleration term $\Delta {\dot f}$  can become more important for lower frequency binaries,  { including well-detached systems  (also less affected by tidal effects).}
Only using the  first derivatives $\dot f$ currently available for some verification binaries, we might examine the potential tertiary effects for the space GW detectors.  {The observational studies for the second  derivative $\ddot f$ will be  continued  for ultra-compact binaries other than HM Cancri. Given the long baselines of the existing data, the known systems such as V407Vul will be  interesting targets 
(see  \citealt{2018MNRAS.480..302K} for a list for LISA's verification binaries).}

It should be also noted  that,
besides the tertiary effects, we might make pilot studies for future GW detectors, by appropriately using their verification binaries.

As omini-directional detector,  free from interstellar absorption, LISA is expected to individually resolve  $\sim 10^4$ Galactic ultra-compact binaries in $\sim 4$ years \citep{2022arXiv220306016A}. It will provide us with not only the orbital phase information for the detected binaries but also their sky positions and inclination  angles.  We will be able to identify many electromagnetic counterparts by  followup observation (e.g., by the help of the predicted eclipse timing).  Then, utilizing archival data of short cadence surveys taken long before LISA's launch  (see e.g., \citealt{2019MNRAS.483.5518K,2022arXiv221214887D}),  we might measure their second derivatives $\ddot f$ with a  time baseline much   longer than the operation period of LISA.

\section{summary}

The interacting binary HM Cancri occupies a central position among the verification binaries for space GW detectors. 
Motivated by the recent development on its long-term orbital analysis, we discussed a potential tertiary perturbation  to this binary.  We found that tertiary effects are likely to have a limited impact on observed first derivative $\dot f$. However,  the magnitude of the reported second derivative $\ddot f\sim -10^{-26}{\rm Hz\,s^{-2}}$ can be easily generated by a dark tertiary  such as an old white dwarf with an outer orbital period of $P\sim250$yr.

\section*{Acknowledgements}
This work is supported by JSPS Kakenhi Grant-in-Aid for  Scientific Research (Nos.~ 17H06358,  19K03870 and 23K03385).

\section*{DATA AVAILABILITY}
The data underlying this article will be shared on reasonable request to the corresponding author.

\bibliographystyle{mn2e}


\end{document}